\documentclass{vgtc}                          

\ifpdf
  \pdfoutput=1\relax                   
  \usepackage{graphicx}                
  \DeclareGraphicsExtensions{.pdf,.png,.jpg,.jpeg} 
\else
  \usepackage{graphicx}                
  \DeclareGraphicsExtensions{.eps}     
\fi%

\graphicspath{{figures/}{pictures/}{images/}{./}} 

\usepackage{microtype}                 
\PassOptionsToPackage{warn}{textcomp}  
\usepackage{textcomp}                  
\usepackage{mathptmx}                  
\usepackage{times}                     
\usepackage{cite}                      
\usepackage{tabu}                      
\usepackage{booktabs}                  
\usepackage{color}

\onlineid{0}

\vgtccategory{Technologies \& Applications }

\vgtcinsertpkg

\title{Virtual Environments for Rehabilitation of Postural Control Dysfunction}

\author{Zhu Wang\thanks{e-mail: zhu.wang@nyu.edu}\\ %
        \scriptsize New York University %
\and Anat Lubetzky\thanks{e-mail: anat@nyu.edu}\\ %
     \scriptsize New York University %
\and Marta Gospodarek\thanks{e-mail: mo1417@nyu.edu}\\ %
          \scriptsize New York University %
\and Makan TaghaviDilamani\thanks{e-mail: mt3299@nyu.edu}\\ %
          \scriptsize New York University %
\and Ken Perlin\thanks{e-mail: ken.perlin@gmail.com}\\ %
     {\scriptsize New York University}}

\abstract{We developed a novel virtual reality [VR] platform with 3-dimensional sounds to help improve sensory integration and visuomotor processing for postural control and fall prevention in individuals with balance problems related to sensory deficits, such as vestibular dysfunction (disease of the inner ear). The system has scenes that simulate scenario-based environments. We can adjust the intensity of the visual and audio stimuli in the virtual scenes by controlling the user interface (UI) settings. A VR headset (HTC Vive or Oculus Rift) delivers stereo display while providing real-time position and orientation of the participants' head. The 3D game-like scenes make participants feel immersed and gradually exposes them to situations that may induce dizziness, anxiety or imbalance in their daily-living. }

\CCScatlist{
  \CCScatTwelve{Human-centered computing}{Visualization}{Visualization design and evaluation methods}{}
}

\begin{document}

\maketitle

\section{Introduction}

According to the U.S. Centers for Disease Control and Prevention, falls are the top public health problem for older Americans. Fall-related injuries bring one to the emergency room every 11 seconds and cause one death every 19 minutes. One in four 65+ Americans experiences at least one fall each year. With that, the direct cost of the medical attention for fall injuries reached 50 billion US Dollars in 2015 \cite{fallFacts}.

Vestibular dysfunction, a disease of the inner ear, is a condition that leads to imbalance and dizziness and significantly increases fall risk \cite{stevens2006costs} \cite{agrawal2009disorders}. The National Health and Nutrition Examination Survey estimated that 35.4 of adults in the United States sought medical attention for a vestibular dysfunction between 2001-2004. Every year, 7 million clinic visits are due to dizziness\cite{sloane1989dizziness}\cite{kroenke2000common}. People with chronic dizziness and imbalance are two to three times more likely to fall in comparison to people without chronic dizziness \cite{ko2006chronic}, with one report indicating a 12-fold increase in the fall risk among individuals with vestibular dysfunction \cite{agrawal2009disorders}. The increased fall risk is \cite{agrawal2009disorders, ko2006chronic}, particularly in complex environments with busy visual and auditory cues. In addition, once a patient experienced loss of balance or dizziness in a certain environment, they are likely to develop anxiety of that situation or avoid it altogether, thus reducing their participation in the community, especially in urban environments.

To help patients improve their balance function and adapt to various sensory stimuli, therapists need to set up complex environments and simulate the stimuli in patients' rehabilitation at the clinics. For that reason, virtual reality (VR) seems to be appropriate in the realm of rehabilitation \cite{whitney2013symptoms}. VR can maximize the advantages of computer technology to render all kinds of semi-realistic 3D models and auditory stimuli that exist in the real world.


Indeed, Physical Therapy has decades of experience of using VR technology, but the use has increased exponentially over the past few years. With the release of modern VR systems (such as HTC Vive and Oculus Rift) since the 2010s, VR technology has become more affordable, portable and with high quality making it a feasible rehabilitation solution for clinics. Compared with conventional balance rehabilitation methods, a VR program can simulate real-life situation without a need to leave the clinic to the actual site (which is often not feasible). A patient can then gradually be exposed to various sensory cues in different contexts with increasing levels of difficulty. All of this is done in a safe and supervised setting, and the patient can `leave' the environment at any time. All the procedures and contents are fully controllable and repeatable which makes large-scale clinical research accessible.

Our collaboration with the clinics gives us an opportunity to build a seamless transfer between new technologies and balance rehabilitation (particularly related to sensory deficits) to improve the quality of life and reduce falls. Our system was developed and refined based on patients' stories and feedback, as well as the needs expressed by therapists. We learned about the type of environments that make patients dizzy, off-balance or anxious (e.g., subway, airports etc.) and the stimuli that enhance those responses (e.g., colors, flow of people etc.). We also interviewed physical therapists about their needs that cannot be met by traditional clinical tools. When designing a rehabilitation software, it is crucial to test its usability in clinics to make sure that clinicians are able to operate the system without technical assistance. Therefore, when the first version of our system was ready, we began descriptive case studies and a clinical implementation study. So far, we have done a usability experiment with 6 physical therapists who treated 17 patients. From those experiments, we gathered user experience when therapists and patients used the system. These data significantly helped us in evolving the functionalities and system designs.



\section{RELATED WORK}

Virtual reality rehabilitation has been shown to be more effective than conventional rehabilitation in regards to physical outcomes for patients with vestibular dysfunction \cite{meldrum2015effectiveness}. It has been suggested that immersive VR environments offer more playability, realistic physical fidelity, and increased cognitive load. With training tasks that mimic the real-world conditions, the transfer from a game-like training program to daily function should be simpler. Patients who participated in VR balance training programs reported they prefer VR over traditional exercises because of more immersive experience, more enjoyment, less after-activity fatigue, and a perception of less difficulty \cite{meldrum2015effectiveness}. For instance, the Computer Assisted Rehabilitation Environment (CAREN) is a virtual reality system for balance assessment and rehabilitation. The CAREN has been tested in a few clinical centers around the US, and demonstrated to be effective in diverse populations \cite{kalron2016effect, sessoms2015improvements}. However, the estimated cost of installing, running and maintaining the CAREN is over \$1 million \cite{kalron2016effect}. Some off-the-shelf augmented reality systems, such as Wii and Kinect, are accessible to rehabilitation but are not as immersive as Head Mounted Displays (HMDs) and the program cannot be customized to individual needs or to create a sensory experience similar to real-life situations \cite{hsu2017three}. Auditory cues have been used as real-time biofeedback \cite{dozza2007auditory}, but are not typically the method of choice for balance rehabilitation of patients with vestibular dysfunction due to the various degrees of hearing loss often accompanying vestibular disorders \cite{sienko2017role}. In our paradigm, we utilize the combination of immersive visual and auditory cues to simulate challenges that cause symptoms to patients in real scenarios. Scenarios are hard to duplicate or repeat in the clinic. In this way, patients will have no risk of safety issues to walk through all environments that may cause them discomfort. In addition, because the environmental stimuli are fully controllable, measurable and reproducible, virtual reality interventions can help clinicians to customize stimuli and target specific impairments within a functional context \cite{morel2015advantages}.

\section{Concept and Significance}

Our portable platform utilizes the HTC Vive or Oculus Rift and includes dynamic environments that capture focused elements of the visual experience to assess the effect of various types of visual cues (e.g., static to dynamic, visual perturbation, lighting, speed of object, etc.) on balance in a safe and immersive manner and without applying destabilizing or nauseating stimuli. The platform targets treatment of multiple domains underlying balance performance including visual dependence, somatosensory contribution, reactive vs. proactive balance, and dual-tasking within a functional context \cite{horak2009balance}.

The system mainly focuses on visual and auditory complexity within a functional context. The functional contexts include different environments in which patients can experience sensory overload. Patients might find themselves unsteady, or light-headed when they are facing a complex environment such as a group of people walking around, cars passing. The complex situations are most likely to occur when people are blending into their social lives, especially when they need to take public transportation such as a subway train or a flight. A subway station or an airport is where patients easily to encounter a massive crowd of people. Yet a closed subway station provides a different sensory experience than a large open airport, hence the functional context is different. Some of our patients avoid taking public transportation (i.e., the subway), others avoid flying.

A fast approaching object can easily make patients lose their balance since people can prepare their body for the situation if they know what kind of gesture and stabilization they need in advance. But in our daily life, it is not possible for people to see everything coming beforehand. For example, it is inevitable to see people throwing balls in parks, people skating and passing by at a fast speed, or even birds flying over your head.

Within different environments, the patients deal with sensory complexity that can be adjusted to their level of symptom, anxiety or imbalance. We can start from an empty environment and progressively increase the number of moving people, different movement directions, speed, etc. In addition, a patient can be trained to react to sudden events, such as a flying ball or an approaching train. This trains their ability to move outside their base of support as well as overcome the fear of responding to a surprising visual event. The task itself depends on patients abilities. Patients can start sitting or standing with support. They can progress to moving their head, looking around, walking or other balance tasks as per the therapist's judgment and the patient's symptoms.

To enhance patients' ability to participate in the community and prevent people from falling, an ideal therapy would be a simulation of the same scenarios where the patients can conquer the stress with no safety issues. Unfortunately, it is not feasible to set up a training session in a real airport or subway station. However, VR can help clinicians to build a simulation of the scenarios which make the patients stressful and anxious without stepping out of clinic rooms. Instead of taking patients physically to the complex environments, VR instantly transforms clinic rooms into any complex environment so that all kinds of precaution and arrangement can ensure the safety of the patients. The adjustable level of content complexity helps clinicians to customize proper training sessions for all types of patients who have different balance disorder situations.

The low cost of our platform components will enable us to reach out to multiple clinics and patients at fall risk that otherwise have limited access to such technology. A gap exists in the availability of ecologically valid but low-cost technology that can be widely available for assessment, and treatment of patients with balance problems and fall risk. Our platform offers a unique solution to this problem.
\section{System Design}

\subsection{Virtual Environments}

\begin{figure}[bt]
\centering
\includegraphics[width=0.45\textwidth]{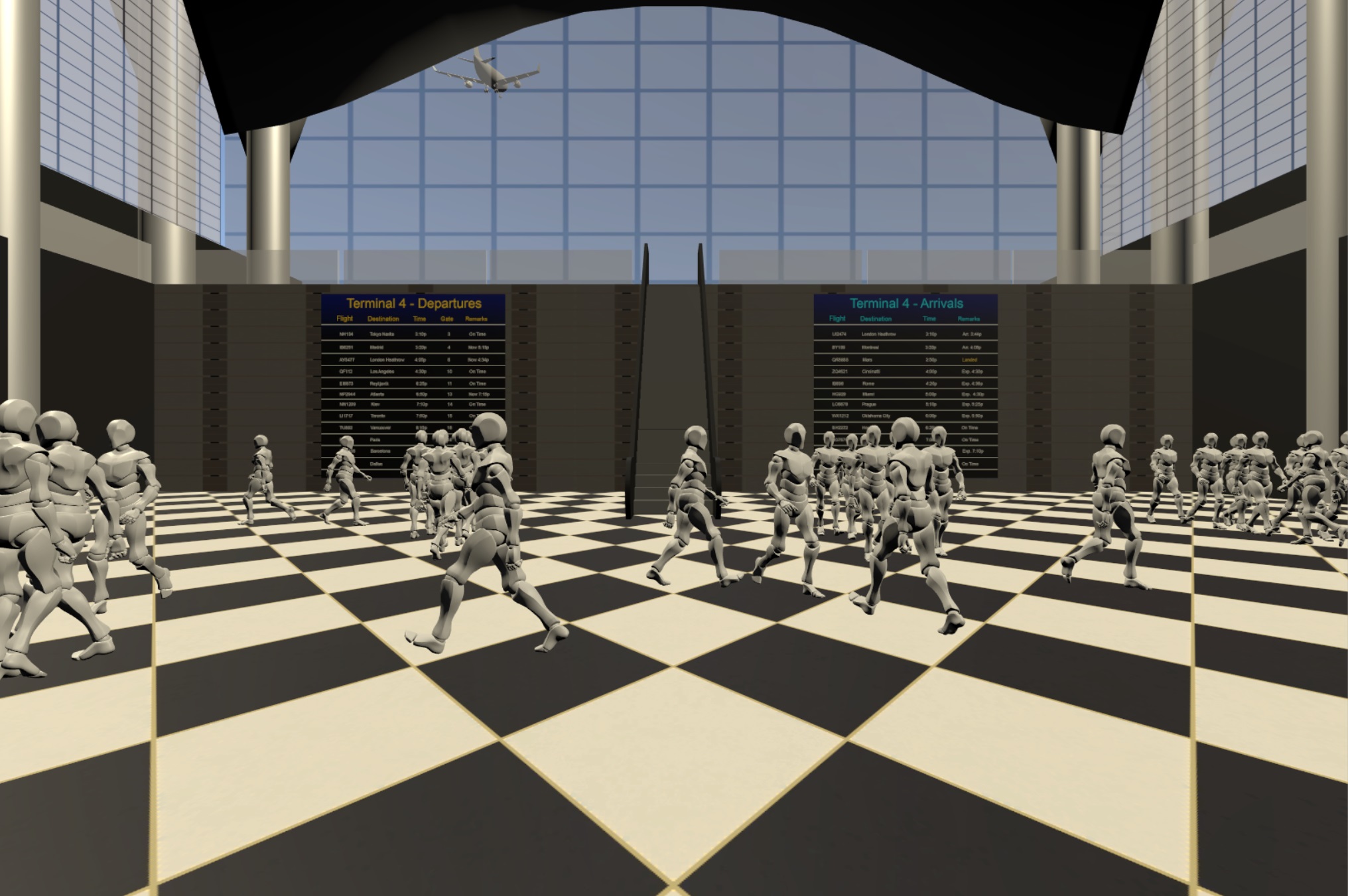}
\caption{Airport scene with high visual intensity (4/4 walking direction level, 4/4 walking amount level).}
\label{fig:airport_scene}
\end{figure}

\begin{figure}[bt]
\centering
\includegraphics[width=0.45\textwidth]{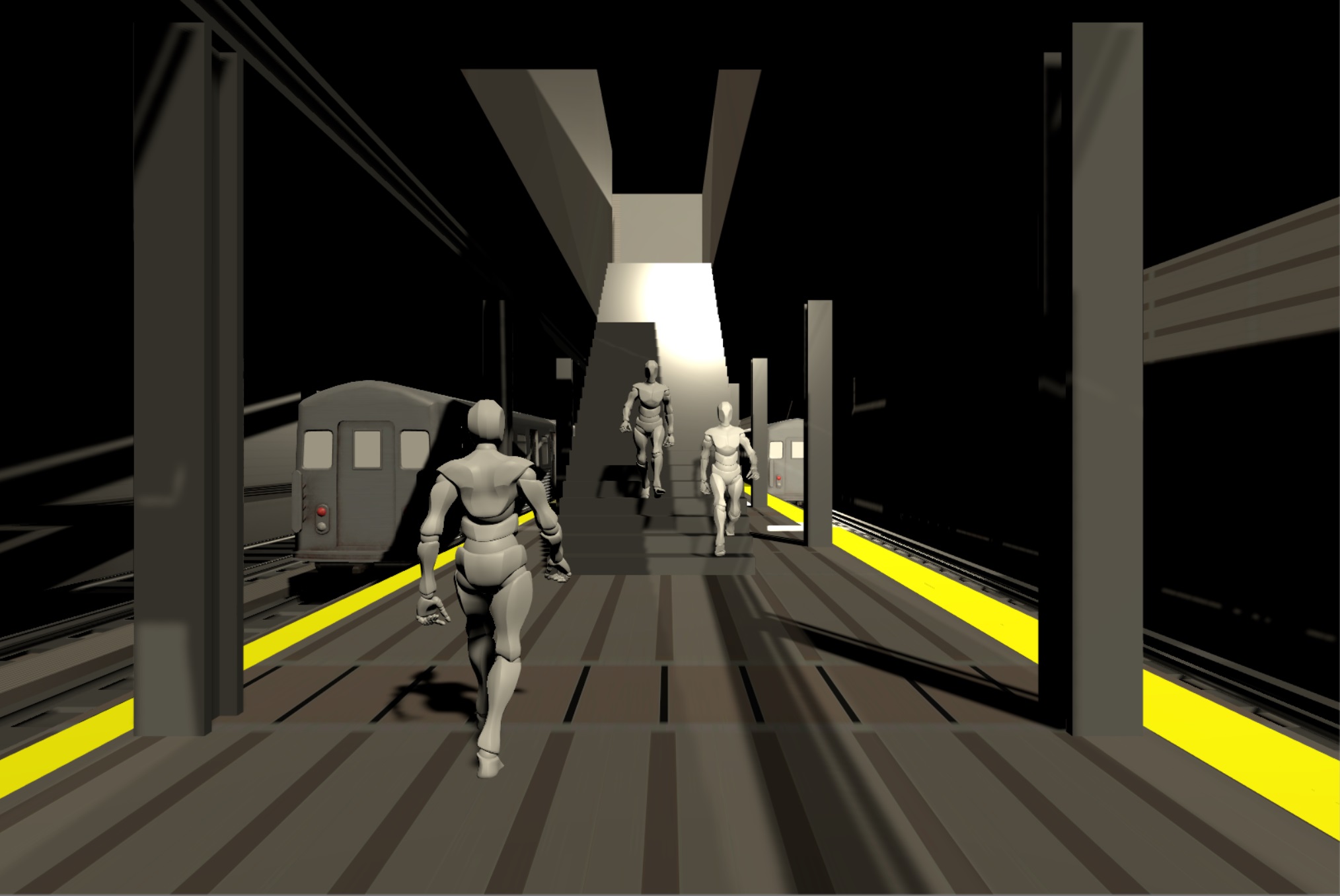}
\caption{Subway scene with low visual intensity (1/4 walking direction level, 1/4 walking amount level).}
\label{fig:subway_scene}
\end{figure}

In our system, there are 4 scenes: airport, subway, city, ball\&park.

The airport scene has a 3D graphics model of an airport terminal which simulates a real airport (see Figure \ref{fig:airport_scene}). 3D models of pedestrians walk around at different speeds, and airplanes fly over the terminal randomly in a range of 50-58 seconds. In addition to the sounds of footsteps from the people passing by and the sounds of planes taking off, the scene has the ambient sound which contains crowd chatter, announcements recorded from a real airport, and noises of airport machines and luggage.

The subway scene is a 3D model of a real subway station(see Figure \ref{fig:subway_scene}). It has the same crowd generation module to create the people walking in groups on the platform, mezzanine, and staircase. Subway trains are active on four rails with each rail has a subway car pass by every 35-50 seconds. Similar to the airport scene, the sound layer consists of the noise of the station, people chatting, announcements recorded from a real subway, as well as the sound of footsteps and the trains passing by.

\begin{figure}[bt]
\centering
\includegraphics[width=0.45\textwidth]{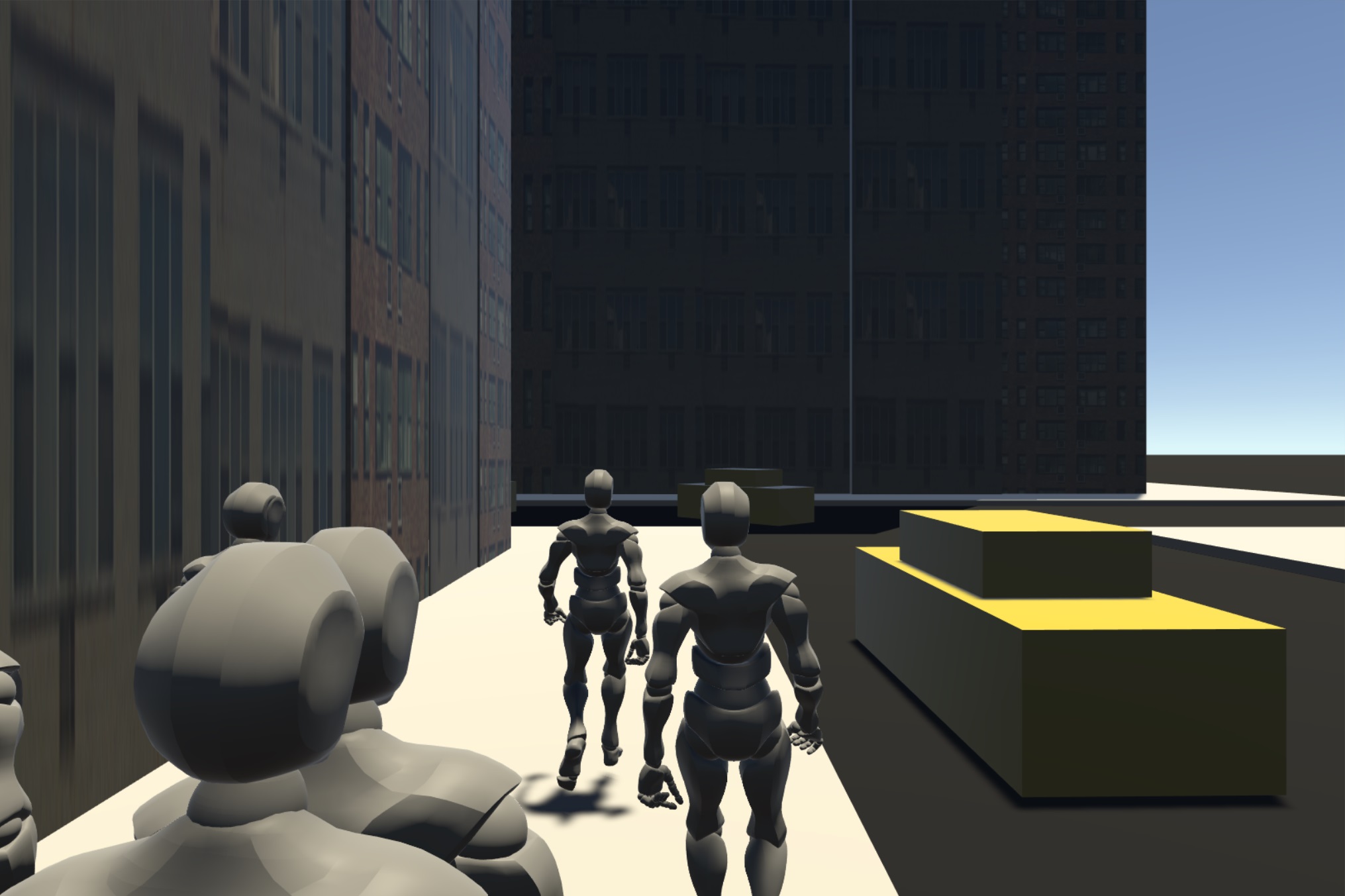}
\caption{City scene with high lighting condition (4/4 brightness) and medium visual intensity (2/4 walking direction level, 2/4 walking amount level, 2/4 car amount level).}
\label{fig:city_scene}
\end{figure}

The city scene simulates 7 blocks of a city which contains vehicles moving around, randomly generated buildings and virtual people on the street (see Figure \ref{fig:city_scene}). When entering the scene, the initial position is in the middle of a sidewalk, and participants are free to walk and stand on the sidewalk or street. The sound in the scene illustrates the footsteps and conversations of pedestrians, cars passing by and honking, as well as distant sounds of the city, i.e. traffic, construction noise, and ambulance sirens.

\begin{figure}[bt]
\centering
\includegraphics[width=0.45\textwidth]{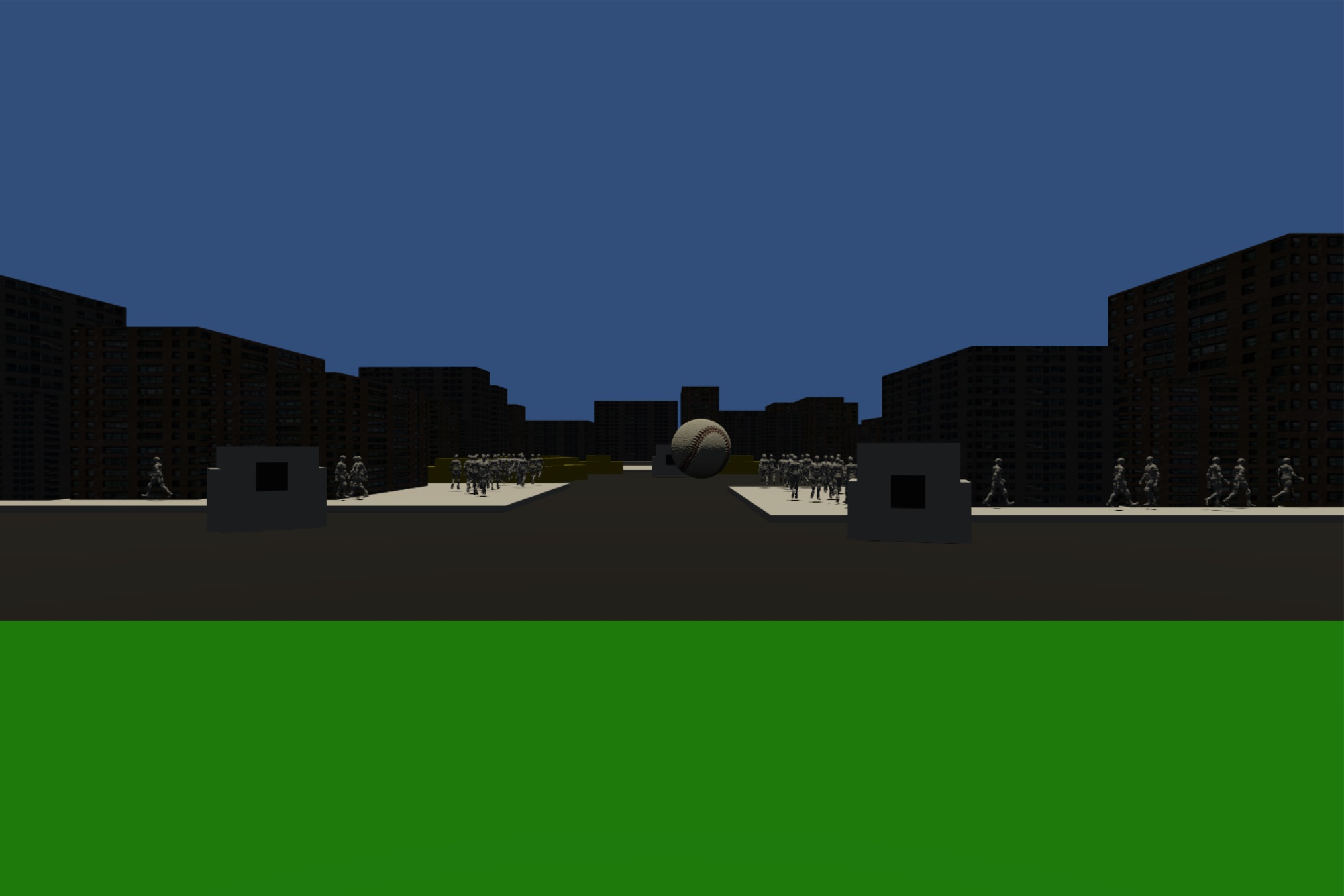}
\caption{Ball\&park scene contains a ball flying near the player's head and with medium lighting condition (2/4 brightness) and high visual intensity (3/4 walking direction level, 3/4 walking amount level, 3/3 car amount level).}
\label{fig:ball_scene}
\end{figure}

The ball \& park scene has a square-shaped park in the middle of a city and tennis ball machines throwing balls towards participants (see Figure \ref{fig:ball_scene}). The participants stand in the middle of the park and need to avoid the tennis balls from 3 tennis ball machines. The therapist can choose how many balls will be activated at the same time (from 0 to 3) and from which direction(s). Once the ball machines are activated, every two balls from each ball machine have a randomized interval between 2-4 seconds. People walk back and forth on the sidewalks, and vehicles run around the street blocks. The player can hear the sound effect of the ball flying near the head, with an addition to the background sounds of the park: footsteps and conversations of pedestrians, birds, and wind in the trees, and distant cars passing by.

Because of the randomized timing, the fast flying ball in the scene is a sudden and unpredictable event which requires patients to move outside of their base of support and may elicit anxiety or loss of balance.

The airport scene, subway scene, and city scene are the simulations of our daily life but with different space types. The airport scene is a big enclosed space, the subway scene is a confined, enclosed space and the city scene is an open space. They can help clinicians to explore the probable effect that the different space types can make on their patients.

Similar to the various sunlight conditions in a real open-air environment, the city and ball\&park have four levels of lighting ranging from 0 to 3. The brightest level simulates the sunlight at noon of a sunny day, the second level provides a cloudy day or morning sunlight, the third level represents dim lighting, and the fourth level represents a dark scene. Clinicians can control the lighting level when they need to create a lighting variant environment for their patients to adapt to different lighting conditions.

\subsection{User Interface and Content Control}

\begin{figure}[bt]
\centering
\includegraphics[width=0.45\textwidth]{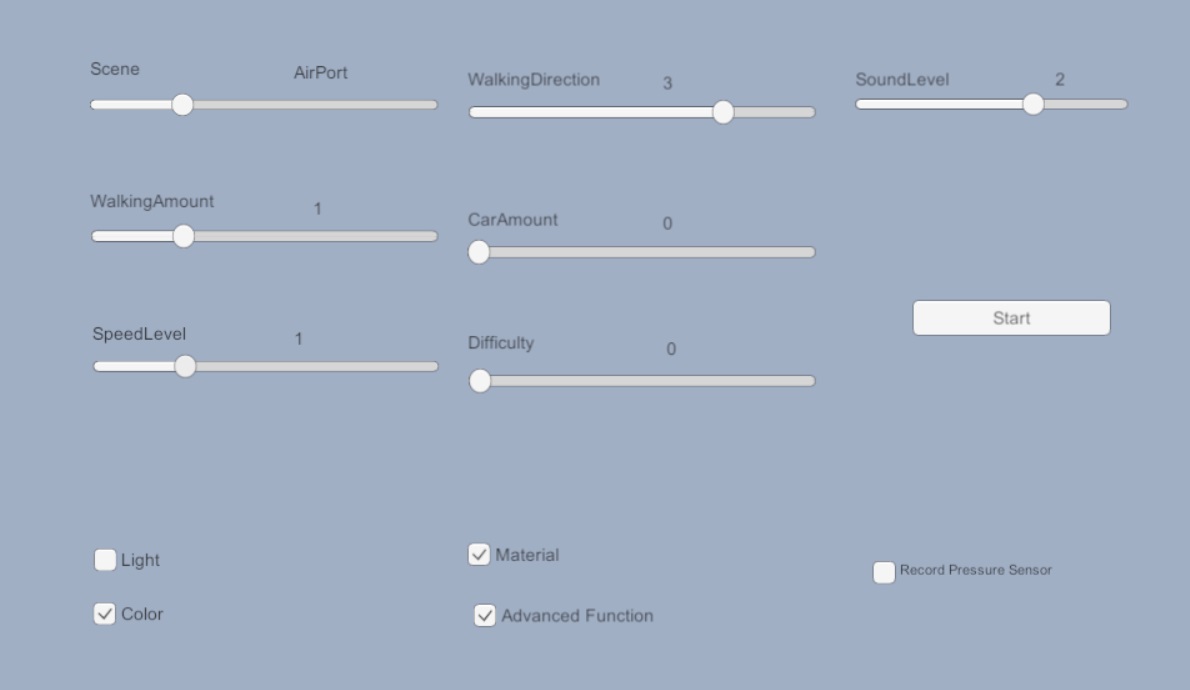}
\caption{The main menu of the platform. The user interface is able to fully control the levels of the visual and auditory stimuli by using the sliders, checkboxes, and buttons.}
\label{fig:ui_design}
\end{figure}

\begin{figure}
\centering
\includegraphics[width=0.5\textwidth]{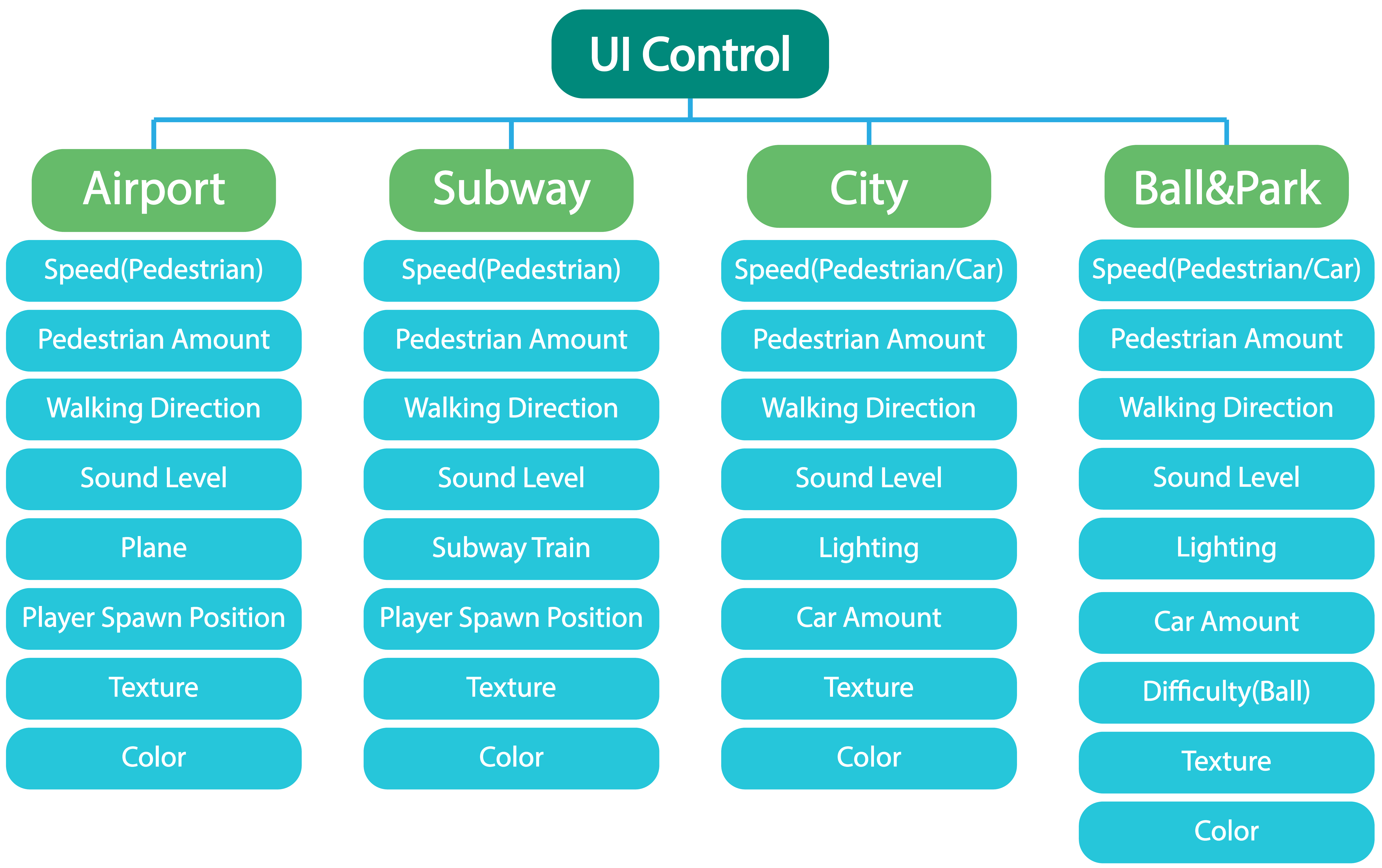}
\caption{UI on the main menu can help users to choose from the four scenes and control the complexity level of the visual and auditory contents in each scene.}
\label{fig:ui_control}
\end{figure}

To make the platform easier to use, we made an intuitive user interface with sliders, checkboxes, and buttons for users to choose scenes, change levels of visual or auditory stimulus, and enable/disable graphics effects or contents (see Figure \ref{fig:ui_design} and Figure \ref{fig:ui_control}).

The ``Scene'' slider helps players to choose between the airport, subway, city, or ball\&park scenes. In the airport, subway, and city scenes, we control the pedestrians' walking speed by modifying ``Speed'', directions and paths of walking pedestrians by ``WalkingDirection'', and quantity in each walking group by ``WalkingAmount''. Choosing 0 speed provides a static scene (no movement), and 1-3 gives them the options of low, medium, and high speed. Additionally, the speed of the cars in the city scene can be controlled individually whereas the speed of the airplane in the airport scene, and the speed of the subway train in the subway scene is fixed.

``SoundLevel'' with range 0-2 controls the complexity of the auditory cues in the scenes. The overall amplitude of the sound layer is self-selected to the highest comfortable level of the user.

``CarAmount'' ranging from 0 to 3 controls the number of cars on the street of the city scene and the ball\&park scene. Zero will generate no cars, and 1-3 are for low, medium and high quantity accordingly. ``Difficulty'' adjusts the difficulty level of the ball\&park scene. In level 0, only one tennis ball machine is activated, and it shoots a ball every 2 to 4 seconds. Level 1-3 each has one of the three ball machines activated randomly, and the interval between two balls ranges from (3 + \#level) to (5 + \#level) seconds. Level 4 has two of the three ball machines activated randomly, and the interval between two balls from each ball machine ranges from 3 to 5 seconds.

The checkboxes help users to choose if they want to enable or disable the details of the secondary graphics contents in the Airport, city and subway scene. ``Light'', ``Color'' and ``Material'' checkboxes respectively control the lighting in the scenes, colors, and materials on the buildings, floors, cars, walls, posters, etc.

\subsection{Sound Implementation}
Previous research indicates that sound can have a significant influence on body balance. Results showed that attributes of sound like frequency \cite{soames1992influence} and movement \cite{park2011effects} can both increase \cite{russolo2002sound}, or decrease\cite{stevens2016auditory} the postural sway depending on the context.

Taking into account that the auditory system has an important role in maintaining body balance, the implementation of sound in our system is based on spatial audio, particularly dynamic binaural rendering on headphones. Spatial audio is widely used in VR applications as it has the advantage of creating a level of realism not possible to achieve in traditional stereo systems \cite{begault20003,chandrasekera2015virtual}. When using binaural audio technology, spatial auditory cues are encoded into a mono audio signal which results in the change of the perceived localization of sound. Using the head-tracking data, audio is modulated according to the position of the listener's head. The technology allows positioning of sound sources in any direction around the listener, thus enabling the creation of soundscape much more similar to reality and more immersive. The use of 3D sound is especially important for this project as the goal of the system is to create a simulation which combines different sensory cues present in real-life situations.

Audio assets used in the system are divided into two main groups: sound objects and ambiances. Sound objects are attached to the visual objects in the scene and their position is changing accordingly. These include the sounds of footsteps, trains, announcements, cars, balls, airplanes, etc. Ambiances are created separately for each scene and they do not change with the position of the listener and objects. These include different background sounds, i.e. sounds of the crowd chatter, distant trains, wind, birds, traffic and general room tone of each of the spaces. To achieve the high level of realism, most of the sounds were gathered from original recordings performed in subway stations and streets. The ambiances were recorded using ambisonics technology. Ambisonics is a full-sphere surround sound format which allows capturing the soundscape from every direction \cite{gerzon1973periphony}. During the playback, the ambisonics sound layer changes depending on the rotation of the listener's head which enables the faithful reproduction of the 360\textdegree environment with the real position of the sound objects.

All of the sounds used in the system are assigned to three different intensity levels which relate to the increasing complexity of the soundscape and amount of auditory stimuli played at the same time from different directions. The lowest level has no sound, the medium level has a minimum number of the sound effects with moderate complexity of background sounds, whereas the highest level creates a bustling auditory environment with a large number of sound cues. Clinicians can adjust the intensity of the audio layer depending on the symptoms and abilities of each patient.

\subsection{Dynamic Collision Detection}

In the early versions, our system did not have a collision detection mechanism. The walking avatars were not aware of the existence of other game objects in the scenes so that they could run through the player or the other walking avatars if their paths got across. This was a little intimidating for some patients and reduced the level of perceived realism for others. For a rich visual project aiming at a certain degree of reality, we desired every avatar to make a detour and plan a new route when they find the player or other avatars on their way so that it would not collide them.

\begin{figure}[bt]
\centering
\includegraphics[width=0.45\textwidth]{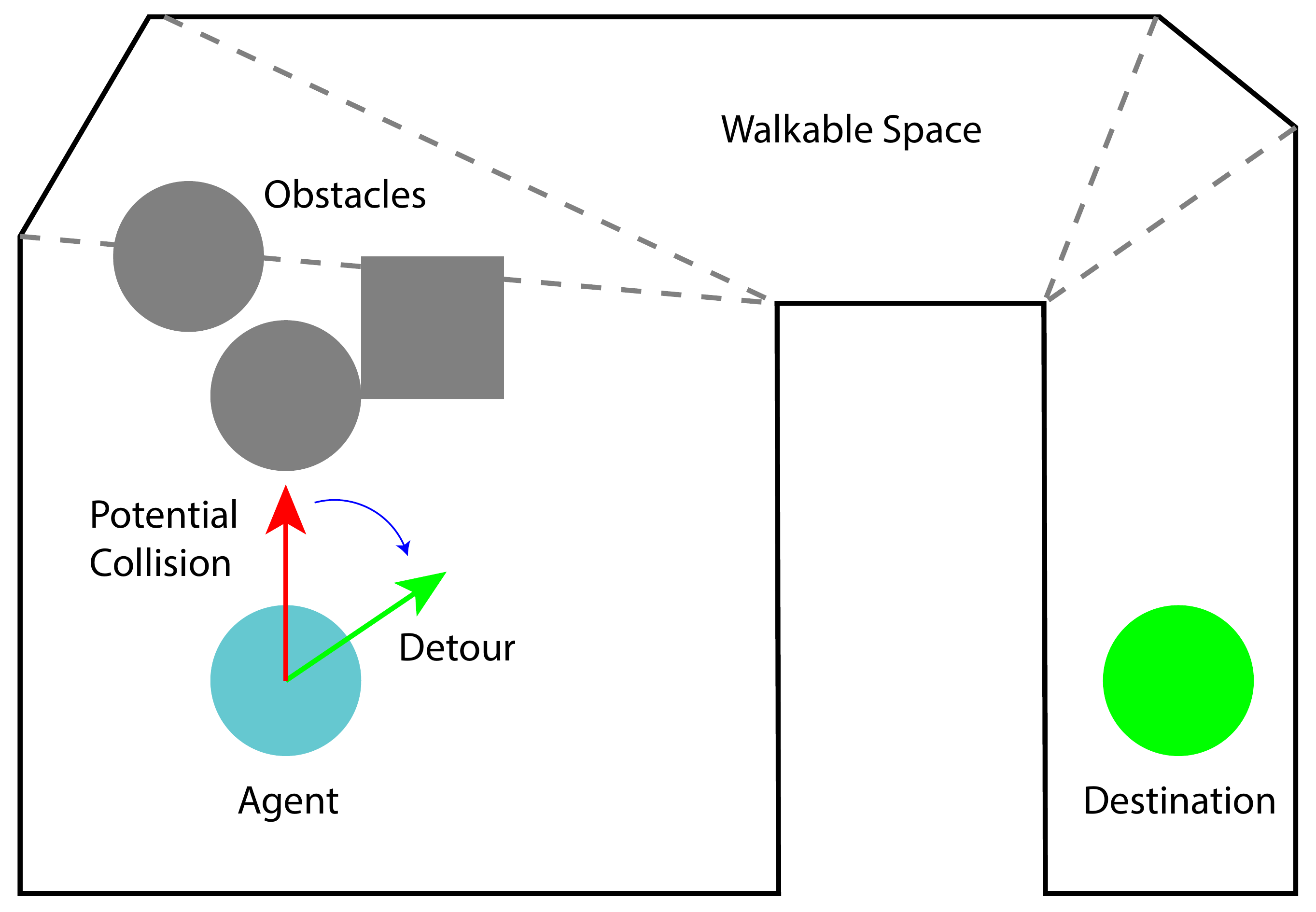}
\caption{NavMesh(Navigation Mesh) defines walkable space by using polygons. An agent makes a detour when it detects potential collision during travel to the destination.}
\label{fig:NavMesh}
\end{figure}

Unity provides a navigation component called NavMesh (Navigation Mesh, see Figure \ref{fig:NavMesh}) which defines the walkable areas by polygons and produces waypoints for game objects to walk through the walkable polygons. The NavMesh consists of two modules: agent and obstacle. The agent module helps to find a walkable path and avoid obstacles, and the obstacle module defines the objects needs to be avoided. In our scenes, the walking avatars should be both agent and obstacle. Because they need to avoid the player and the other walking avatars, and they need to be avoided by the others in the meantime. However, the two modules cannot be activated simultaneously to the same game object according to Unity's official document and our experiments. If we activate the two module on a game object at the same time, the object will try to avoid itself so it would jitter or walk in circles. To have it work properly, we still add two modules to the avatars, but alternatively turn on only one of them at a time. Because the pathfinding is only needed when a collision is coming, we activate the agent and inactivate obstacle module only when a potential collision is detected; After the new waypoints are generated, the avatar enables the obstacle and disables the agent module for the next potential collision detection.

\subsection{Hardware Setup}

\begin{figure}[bt]
\centering
\includegraphics[width=0.45\textwidth]{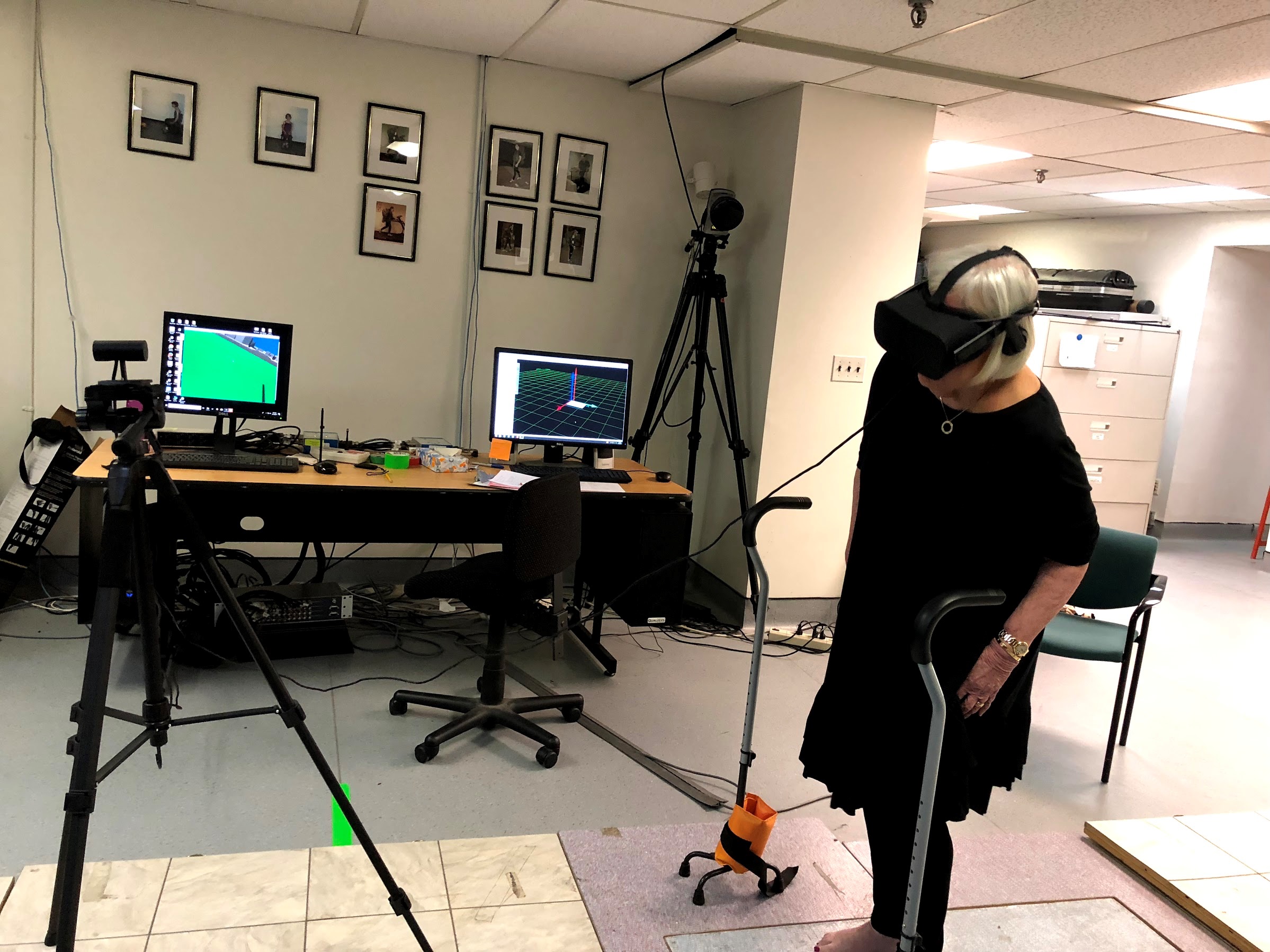}
\caption{An experimental setup. One of our case-study participants is wearing Oculus Rift and avoiding a flying ball in the park scene.}
\end{figure}

Our system is implemented in C\# language using Unity Engine version 2018.2.0f1(64-bit)(\textcopyright Unity Technologies, San Francisco, California). It uses OpenVR as the API to get full compatibility with all major VR display platforms such as Oculus Rift and HTC Vive. HTC Vive and Oculus Rift both have a resolution of 1080x1200 for each eye, a 90 Hz refresh rate, and 110$^{\circ}$ field of view \cite{viveSpec,oculusSpec}. The audio is implemented using Wwise\textregistered middleware and Google Resonance plugin for 3D audio rendering.

HTC VIVE minimum specifications \cite{viveSpec} are Intel\textregistered Core\texttrademark i5-4590 or AMD FX\texttrademark 8350 for CPU, 4GB for RAM, NVIDIA\textregistered GeForce\textregistered GTX 1070/Quadro P5000 or AMD Radeon\texttrademark Vega 56 for GPU, and Windows\textregistered 8.1 for operation system. The minimum requirements of Oculus Rift \cite{oculusSpec} are Intel i3-6100 or AMD FX4350/Ryzen 3 1200 for CPU, 8GB for RAM, NVIDIA GTX 1050Ti or AMD Radeon RX 470 for GPU (alternatively, NVIDIA GTX 960 or AMD Radeon R9 290), and Windows\textregistered 10 for OS. The sound playback requires closed-back stereo headphones. Otherwise, it is possible to use built-in headphones from Oculus Rift although they provide less isolation from outside environment.

Our lab setup is an Alienware\textregistered laptop 15 R3 running Windows\textregistered 10 with 8GB RAM, Intel i7-7820HK CPU, Nvidia\textregistered GTX 1080 Max-Q GPU, and Bose SoundTrue\textregistered around-ear headphones II. During our test, the platform runs on the highest level of graphics contents can work at over 120 fps with either HTC Vive or Oculus Rift.

%


\section{Experiments}

\subsection{Usability Experiment}

%


For a new technology to be implemented successfully in clinical practice, users should be involved in early stages of development and evaluation of the technology \cite{van2011ambient, shah2009developing}. The International Organization for Standardization (ISO,9241-11) defined the usability of a device as the extent to which the device can be used by specific users to achieve specific goals \cite{din19989241}.

For over 4 months, the experiments enrolled 6 therapists and 17 patients with peripheral or central vestibular dysfunction. 4 of the patients did not finish the whole experiment due to medical reasons, scheduling problems or insurmountable anxiety.

\subsubsection{Airport}

Clinicians control the walking directions of the virtual pedestrians, and they can choose the initial spawn position for participants from the hall, the second floor, or the stairs. They can enable or disable the taking-off planes, the posters and the walls, the texture on the floor, etc. Clinicians also can change the sound level between no sound, low sound and high sound depending on the capabilities of patients.

This scene was developed based on patients' stories, such as:
\begin{itemize}
  \item ``I am able to travel but in busy airports I sit in a wheelchair... people coming from front and back and all the sounds....''
  \item ``when I walked into the airport lobby and there was a patterned floor. I almost fell over.''
\end{itemize}

\subsubsection{Subway}

The intervention includes manipulations on the quantity and walking directions of the virtual pedestrians; choose the initial spawn position for participants from the platform, the mezzanine, or the stairs; change the audio intensity level between no sound, low and high intensity sound. Clinicians can enable or disable passing trains and pedestrians, the textures and colors on the floor, pillars, walls, etc.

The subway scene was developed based on patients' stories, such as:
\begin{itemize}
  \item ``I don't feel comfortable standing on the platform, I feel like I will fall over.''
  \item ``I don't take the subway, there are too many people and I am afraid of falling onto the track if someone bumps into me.''
\end{itemize}

\subsubsection{City}

Clinicians can manipulate the quantity, speed, walking directions of the virtual pedestrian; the quantity and speed of cars on the street; enable/disable cars and pedestrians; the textures and colors of the buildings and vehicles; 4 levels of lighting condition; audio intensity level between no sound, low sound, and high intensity sound.

Some feedback from patients:
\begin{itemize}
  \item ``This feels like a mild version of the outside experience.''
  \item ``I feel uneasy with the rapid change of light to dark...it's like in the cinema.''
\end{itemize}

\subsubsection{Ball\&park}

The participant needs to avoid 1 or 2 balls approaching their head in each round. Clinicians can manipulate the quantity, speed, walking directions of the virtual pedestrian; the difficulty levels of throwing balls; the quantity and speed of cars on the street; enable/disable cars; the textures and colors of the buildings and vehicles; 4 levels of lighting condition; audio intensity level between no sound, low and high intensity sound.

Some feedback from  patients:
\begin{itemize}
  \item ``If I dodged like that in open space, I would feel dizzy, but I had no problem doing it within the scene!''
  \item ``I think that would be a great therapy for making me move my head...If you played with where it goes''
\end{itemize}

Patients completed the Short Feedback Questionnaire (SFQ) \cite{erez2013comparing} after the first exposure to a scene. The SFQ is designed to obtain information about the subjective responses of the participants to the VR experience in each scenario, including enjoyment, perceived difficulty of the task, sense of presence and side effects. Five physical therapists completed the System Usability Scale once per month \cite{zare2018developing}. The Usability questionnaire includes ten items which provide a global view of subjective assessment of a system's usability. The therapists also had a monthly meeting with the Principal Investigator to discuss feedback regarding the system. Average usability scores were 70 when PT used the system < 5 times, dropped to 63 at 5-10 times usage and increased to 73 above 10 times (ideal is 100, 68 is acceptable). SFQ scores per scene were 16 for the subway, airport, city and 15 for ball\&park (20 is the best score). Enjoyment and `feeling inside the task' were rated above 4 / 5 for all (highest on the subway), and 3.75 for ball\&park. The City scene was the most commonly used scene. Discomfort and difficulty (scored 1-5 where lower is less discomfort) were around 1 for the airport and park, and close to 2 for the subway and city. In the majority of the patients, there was no increase in symptoms from baseline to post VR session.

\subsection{Case Studies}

In two in-depth case studies, two female patients aged 72, 75 with unilateral peripheral vestibular hypofunction participated in 8 individualized sessions. Each session was 45 minutes long, averaging about 5-6 scenes per session for 1-4 minutes. Typically, sessions were performed once every week. The complexity of the environments and the level of stimulus intensity were gradually increased depending on the patients' progress and symptoms. Participants were asked to walk, turn around, sidestep and turn their head within the VR environment.


We used a combination of patient-reported outcomes and functional tests of gait and balance to assess patients' progress. The Functional Gait Analysis \cite{marchetti2014responsiveness} contains 10 walking tasks (e.g., step over an obstacle, walk and turn, stair climbing) to assess dynamic balance and postural stability during gait. The maximal score is 30, and a score under 22 is associated with high fall risk in older adults. The Four Step Square test \cite{whitney2007reliability} accesses the ability to step over objects forward, sideways, and backward as fast as possible. The Activities-Specific Balance Confidence is a questionnaire to report one's level of confidence that he/she will not lose their balance during various activities of daily living (0 means no confidence in their balance, 100\% is full confidence, a score under 67\% relates to high fall risk \cite{whitney1999activities}). The Visual Vertigo Analog Scale asks the patient to rate the intensity of their dizziness in nine situations of visual motions \cite{dannenbaum2011visual}. The State and Trait Anxiety questionnaire (STAI S / T) \cite{julian2011measures} evaluates participant's anxiety levels in their daily living (Trait) and right now (State). Lower scores are better, and 20 is the lowest possible score.



\begin{figure}
  \begin{minipage}[t][][b]{0.23\textwidth}
    \includegraphics[width=\textwidth]{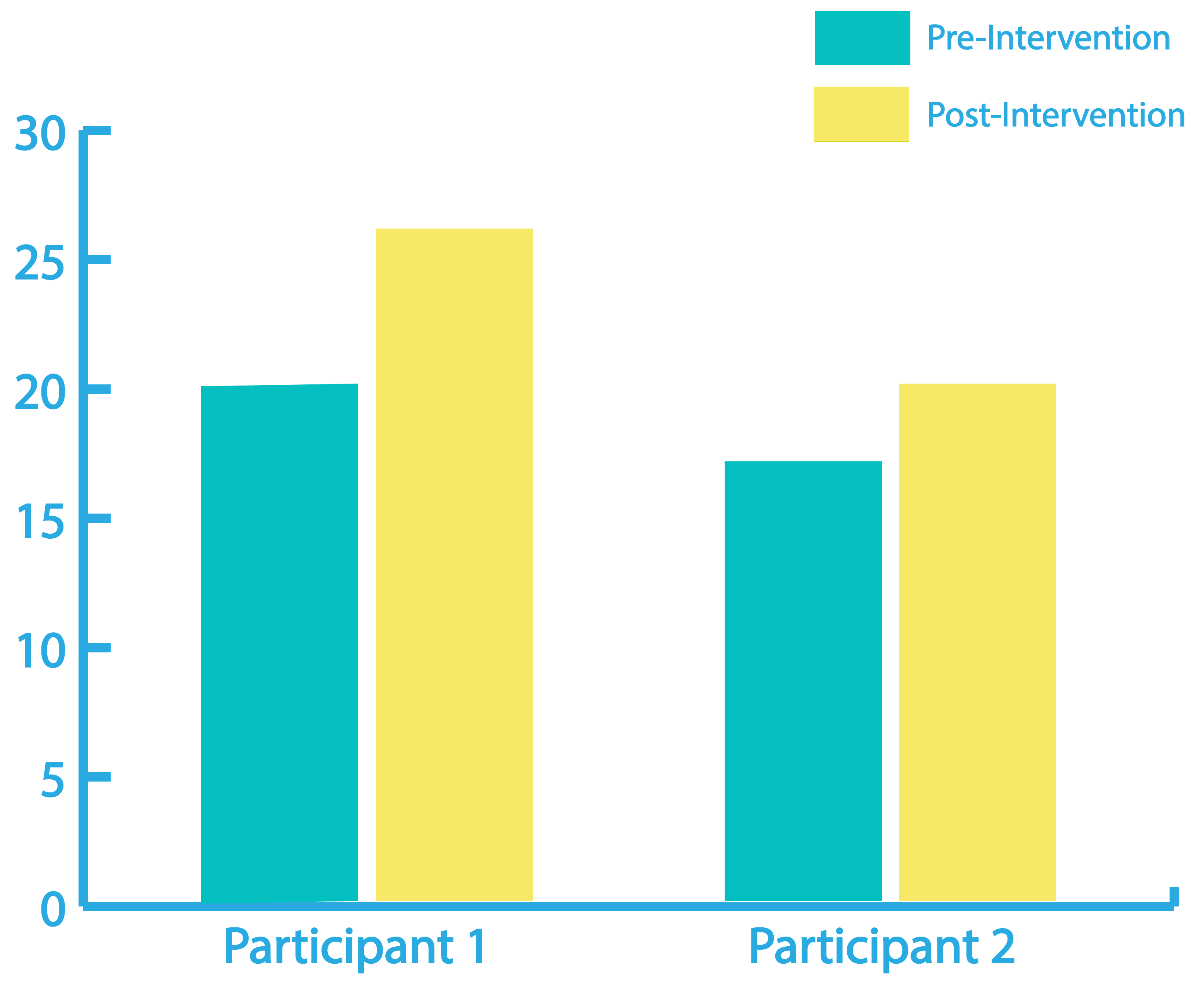}
    \caption{Pre-intervention and post-intervention results of the Functional Gait Analysis of the two participants.}
    \label{fig:fga}
  \end{minipage}
  ~
  \begin{minipage}[t][][b]{0.23\textwidth}
    \includegraphics[width=\textwidth]{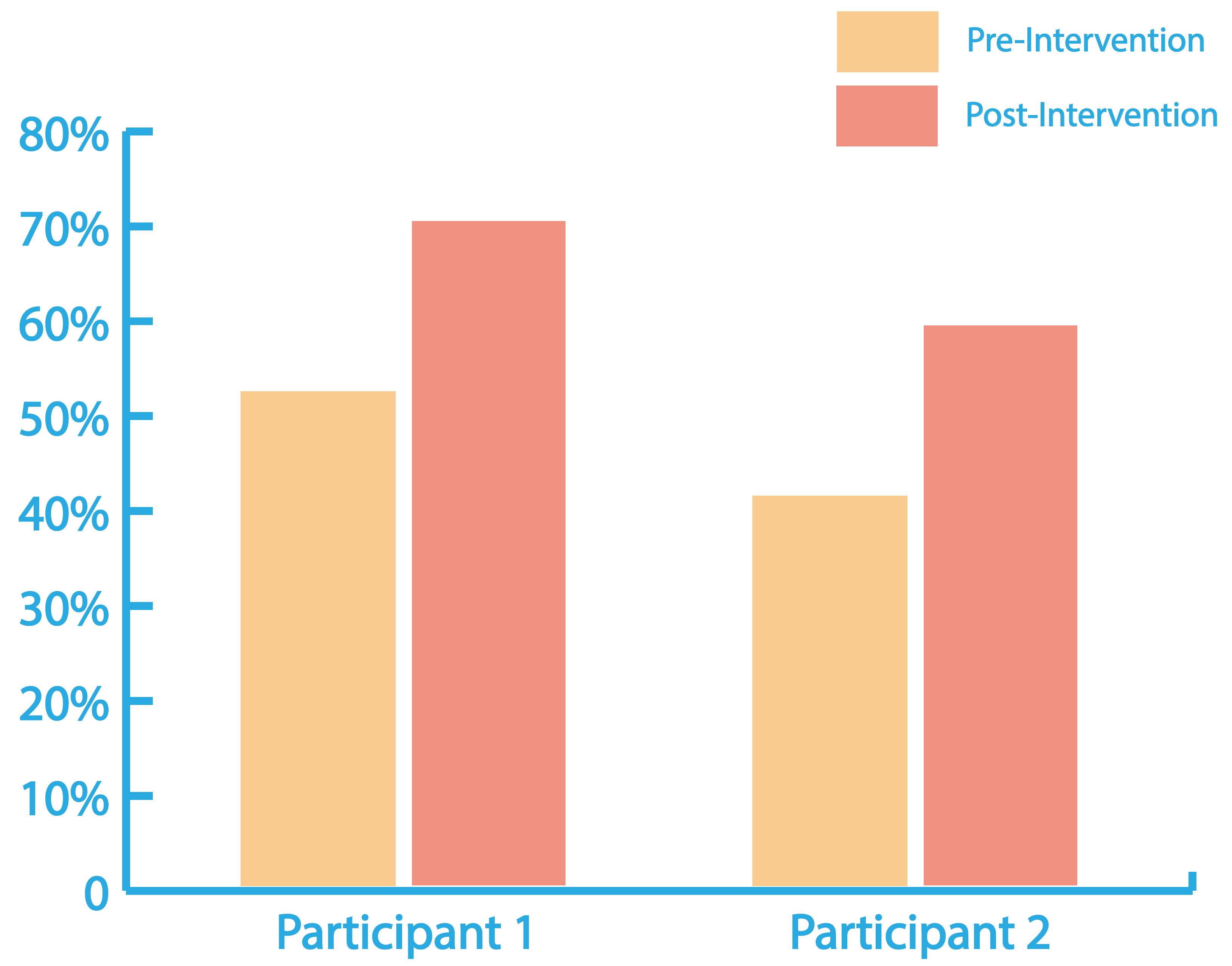}
    \caption{Pre-intervention and post-intervention results of the Activities-Specific Balance Confidence study of the two participants.}
    \label{fig:balance_confidence}
  \end{minipage}
\end{figure}

From baseline to post-intervention, Functional Gait Analysis improved from 20 to 26 (patient 1) and 17 to 20 (patient 2) (see Figure \ref{fig:fga}). Four Step Square Test has not changed for patient 1 (9 seconds) but improved from 15 to 8 seconds in patient 2. Both patients improved their Activities-Specific Balance Confidence (52\% to 70\% and 41\% to 59\%) (see Figure \ref{fig:balance_confidence}), their Visual Vertigo Analog Scale (284mm to 211mm and 640mm to 234mm) and their State (S) / Trait (T) Anxiety (from 46(S)/37(T) to 22(S)/27(T) and from 32(S)/25(T) to 21(S)/21(T)).

\section{Contributions and Limitations}

%
%


Virtual reality (VR) programs have been used for many years in balance rehabilitation, and overall have been shown to be effective for various vestibular disorders\cite{howard2017meta}. Recent advances in head-mounted display (HMD) technology allow for high level of immersion at low costs and minimal technical requirement \cite{lubetzky2017assessment, lubetzky2018feasibility, micarelli2017three}. Our Vive app enables patients to engage in safe yet challenging environments that are not easily reproducible in traditional rehabilitation. Using this platform could improve balance and reduce anxiety in patients with vestibular disorders, especially in hectic environments. This was observed in our case studies and will be further investigated in a clinical trial.

Below are a couple of quotes from the patients participating in the case studies:
\begin{itemize}
  \item ``I love the mezzanine in the airport because I can face my biggest fear of transparent heights but I know that it's virtual and I'm not going to fall''
  \item ``I was able to carry 3 bags on the subway...this has helped me tremendously...''
  \item ``I feel better after doing this then when I walked in."
  \item ``I'm more energized after doing it."
  \item ``Every time I come, I feel better after."
\end{itemize}

In the clinic, our system was found to be feasible, and for the most part, easy to operate. The patients enjoyed the scenes, and the physical therapists confirmed that the system added a dimension to sensory integration rehabilitation that they did not have before. Symptoms were minimal, particularly on the milder scenes. A few implementation challenges should be noted. These include: guarding the patients without blocking the Vive light-houses (if a therapist blocked the light-house, a system restart might be required which takes away treatment time), frequent updates of the PC which may take away the time a therapist has with the patient and then the therapist may choose not to use the system altogether, patients stepping on the cable when walking around, and the cable getting loose. To alleviate these last two limitations, we are now working on transitioning from the current tethered VR system to a wireless setup by using the HTC Vive wireless adapter. In September 2017, HTC announced their wireless solution for VIVE and VIVE pro. The solution uses Intel's highspeed cable-free WiGig network to transmit data between a computer and a VIVE headset. It requires four additional hardware: PCIe WiGig card, wireless link box, Vive Wireless Adapter, and a power bank.  The wireless link box connects to the PCIe card, wirelessly pairs with the Vive Wireless Adapter, and helps to build a fast, low latency wireless network between a computer and the headset. The power bank provides the necessary power to the adapter and the headset. The wireless setup demands the extra weight of the wireless adapter and the power bank, and the WiGig network is not robust if the wireless adapter is further than 20ft from the wireless link box, but patients can have a much better immersive experience when entering the VR world untethered without the constraints of the cable connecting the headset and computer.

Our long-term goal is to provide a streamlined, individualized, user-friendly intervention platform ready to be tested in clinical trials; that will lead to improved participation restriction and quality of life following a customized virtual environments training.

\bibliographystyle{abbrv-doi}

\bibliography{sample-bibliography}

\begin{thebibliography}{10}

\bibitem{agrawal2009disorders}
Y.~Agrawal, J.~P. Carey, C.~C. Della~Santina, M.~C. Schubert, and L.~B. Minor.
\newblock Disorders of balance and vestibular function in us adults: data from
  the national health and nutrition examination survey, 2001-2004.
\newblock {\em Archives of internal medicine}, 169(10):938--944, 2009.

\bibitem{begault20003}
D.~R. Begault and L.~J. Trejo.
\newblock 3-d sound for virtual reality and multimedia.
\newblock 2000.

\bibitem{fallFacts}
CDC.
\newblock Falls among older adults: An overview - home and recreational safety,
  2017.

\bibitem{chandrasekera2015virtual}
T.~Chandrasekera, S.-Y. Yoon, and N.~D'Souza.
\newblock Virtual environments with soundscapes: a study on immersion and
  effects of spatial abilities.
\newblock {\em Environment and Planning B: Planning and Design},
  42(6):1003--1019, 2015.

\bibitem{dannenbaum2011visual}
E.~Dannenbaum, G.~Chilingaryan, and J.~Fung.
\newblock Visual vertigo analogue scale: an assessment questionnaire for visual
  vertigo.
\newblock {\em Journal of Vestibular Research}, 21(3):153--159, 2011.

\bibitem{din19989241}
E.~Din.
\newblock 9241-11. ergonomic requirements for office work with visual display
  terminals (vdts)--part 11: Guidance on usability.
\newblock {\em International Organization for Standardization}, 1998.

\bibitem{dozza2007auditory}
M.~Dozza, F.~B. Horak, and L.~Chiari.
\newblock Auditory biofeedback substitutes for loss of sensory information in
  maintaining stance.
\newblock {\em Experimental brain research}, 178(1):37--48, 2007.

\bibitem{erez2013comparing}
N.~Erez, P.~L. Weiss, R.~Kizony, and D.~Rand.
\newblock Comparing performance within a virtual supermarket of children with
  traumatic brain injury to typically developing children: A pilot study.
\newblock {\em OTJR: occupation, participation and health}, 33(4):218--227,
  2013.

\bibitem{gerzon1973periphony}
M.~A. Gerzon.
\newblock Periphony: With-height sound reproduction.
\newblock {\em Journal of the Audio Engineering Society}, 21(1):2--10, 1973.

\bibitem{horak2009balance}
F.~B. Horak, D.~M. Wrisley, and J.~Frank.
\newblock The balance evaluation systems test (bestest) to differentiate
  balance deficits.
\newblock {\em Physical therapy}, 89(5):484--498, 2009.

\bibitem{howard2017meta}
M.~C. Howard.
\newblock A meta-analysis and systematic literature review of virtual reality
  rehabilitation programs.
\newblock {\em Computers in Human Behavior}, 70:317--327, 2017.

\bibitem{hsu2017three}
S.-Y. Hsu, T.-Y. Fang, S.-C. Yeh, M.-C. Su, P.-C. Wang, and V.~Y. Wang.
\newblock Three-dimensional, virtual reality vestibular rehabilitation for
  chronic imbalance problem caused by m{\'e}ni{\`e}re’s disease: a pilot
  study.
\newblock {\em Disability and rehabilitation}, 39(16):1601--1606, 2017.

\bibitem{viveSpec}
Specifications of headset and recommended minimum computer, 2018.
\newblock https://www.vive.com/us/product/vive-virtual-reality-system/.

\bibitem{julian2011measures}
L.~J. Julian.
\newblock Measures of anxiety: State-trait anxiety inventory (stai), beck
  anxiety inventory (bai), and hospital anxiety and depression scale-anxiety
  (hads-a).
\newblock {\em Arthritis care \& research}, 63(S11):S467--S472, 2011.

\bibitem{kalron2016effect}
A.~Kalron, I.~Fonkatz, L.~Frid, H.~Baransi, and A.~Achiron.
\newblock The effect of balance training on postural control in people with
  multiple sclerosis using the caren virtual reality system: a pilot randomized
  controlled trial.
\newblock {\em Journal of neuroengineering and rehabilitation}, 13(1):13, 2016.

\bibitem{ko2006chronic}
C.~Ko, H.~Hoffman, and D.~Sklare.
\newblock Chronic imbalance or dizziness and falling: Results from the 1994
  disability supplement to the national health interview survey and the second
  supplement on aging study.
\newblock In {\em Annual Meeting of the Association for Research in
  Otolaryngology}, vol.~6, 2006.

\bibitem{kroenke2000common}
K.~Kroenke, R.~M. Hoffman, and D.~Einstadter.
\newblock How common are various causes of dizziness? a critical review.
\newblock {\em Southern medical journal}, 93(2):160--7, 2000.

\bibitem{lubetzky2017assessment}
A.~V. Lubetzky, D.~Harel, H.~Darmanin, and K.~Perlin.
\newblock Assessment via the oculus of visual “weighting” and
  “reweighting” in young adults.
\newblock {\em Motor control}, 21(4):468--482, 2017.

\bibitem{lubetzky2018feasibility}
A.~V. Lubetzky, E.~E. Kary, D.~Harel, B.~Hujsak, and K.~Perlin.
\newblock Feasibility and reliability of a virtual reality oculus platform to
  measure sensory integration for postural control in young adults.
\newblock {\em Physiotherapy theory and practice}, pp. 1--16, 2018.

\bibitem{marchetti2014responsiveness}
G.~F. Marchetti, C.-C. Lin, A.~Alghadir, and S.~L. Whitney.
\newblock Responsiveness and minimal detectable change of the dynamic gait
  index and functional gait index in persons with balance and vestibular
  disorders.
\newblock {\em Journal of Neurologic Physical Therapy}, 38(2):119--124, 2014.

\bibitem{meldrum2015effectiveness}
D.~Meldrum, S.~Herdman, R.~Vance, D.~Murray, K.~Malone, D.~Duffy, A.~Glennon,
  and R.~McConn-Walsh.
\newblock Effectiveness of conventional versus virtual reality--based balance
  exercises in vestibular rehabilitation for unilateral peripheral vestibular
  loss: Results of a randomized controlled trial.
\newblock {\em Archives of physical medicine and rehabilitation},
  96(7):1319--1328, 2015.

\bibitem{micarelli2017three}
A.~Micarelli, A.~Viziano, I.~Augimeri, D.~Micarelli, and M.~Alessandrini.
\newblock Three-dimensional head-mounted gaming task procedure maximizes
  effects of vestibular rehabilitation in unilateral vestibular hypofunction: a
  randomized controlled pilot trial.
\newblock {\em International Journal of Rehabilitation Research},
  40(4):325--332, 2017.

\bibitem{morel2015advantages}
M.~Morel, B.~Bideau, J.~Lardy, and R.~Kulpa.
\newblock Advantages and limitations of virtual reality for balance assessment
  and rehabilitation.
\newblock {\em Neurophysiologie Clinique/Clinical Neurophysiology},
  45(4-5):315--326, 2015.

\bibitem{oculusSpec}
Specifications of headset and recommended minimum computer, 2018.
\newblock https://www.oculus.com/rift.

\bibitem{park2011effects}
S.~H. Park, K.~Lee, T.~Lockhart, and S.~Kim.
\newblock Effects of sound on postural stability during quiet standing.
\newblock {\em Journal of neuroengineering and rehabilitation}, 8(1):67, 2011.

\bibitem{russolo2002sound}
M.~Russolo.
\newblock Sound-evoked postural responses in normal subjects.
\newblock {\em Acta oto-laryngologica}, 122(1):21--27, 2002.

\bibitem{sessoms2015improvements}
P.~H. Sessoms, K.~R. Gottshall, J.-D. Collins, A.~E. Markham, K.~A. Service,
  and S.~A. Reini.
\newblock Improvements in gait speed and weight shift of persons with traumatic
  brain injury and vestibular dysfunction using a virtual reality
  computer-assisted rehabilitation environment.
\newblock {\em Military medicine}, 180(suppl\_3):143--149, 2015.

\bibitem{shah2009developing}
S.~G.~S. Shah, I.~Robinson, and S.~AlShawi.
\newblock Developing medical device technologies from users' perspectives: a
  theoretical framework for involving users in the development process.
\newblock {\em International journal of technology assessment in health care},
  25(4):514--521, 2009.

\bibitem{sienko2017role}
K.~Sienko, S.~Whitney, W.~Carender, and C.~Wall~III.
\newblock The role of sensory augmentation for people with vestibular deficits:
  real-time balance aid and/or rehabilitation device?
\newblock {\em Journal of Vestibular Research}, 27(1):63--76, 2017.

\bibitem{sloane1989dizziness}
P.~D. Sloane.
\newblock Dizziness in primary care: results from the national ambulatory
  medical care survey.
\newblock {\em Journal of Family Practice}, 29(1):33--39, 1989.

\bibitem{soames1992influence}
R.~Soames and S.~Raper.
\newblock The influence of moving auditory fields on postural sway behaviour in
  man.
\newblock {\em European journal of applied physiology and occupational
  physiology}, 65(3):241--245, 1992.

\bibitem{stevens2006costs}
J.~A. Stevens, P.~S. Corso, E.~A. Finkelstein, and T.~R. Miller.
\newblock The costs of fatal and non-fatal falls among older adults.
\newblock {\em Injury prevention}, 12(5):290--295, 2006.

\bibitem{stevens2016auditory}
M.~N. Stevens, D.~L. Barbour, M.~P. Gronski, and T.~E. Hullar.
\newblock Auditory contributions to maintaining balance.
\newblock {\em Journal of Vestibular Research}, 26(5-6):433--438, 2016.

\bibitem{van2011ambient}
J.~Van~Hoof, E.~J. Wouters, H.~R. Marston, B.~Vanrumste, and R.~Overdiep.
\newblock Ambient assisted living and care in the netherlands: the voice of the
  user.
\newblock {\em International Journal of Ambient Computing and Intelligence
  (IJACI)}, 3(4):25--40, 2011.

\bibitem{whitney1999activities}
S.~Whitney, M.~Hudak, and G.~Marchetti.
\newblock The activities-specific balance confidence scale and the dizziness
  handicap inventory: a comparison.
\newblock {\em Journal of vestibular research}, 9(4):253--259, 1999.

\bibitem{whitney2007reliability}
S.~L. Whitney, G.~F. Marchetti, L.~O. Morris, and P.~J. Sparto.
\newblock The reliability and validity of the four square step test for people
  with balance deficits secondary to a vestibular disorder.
\newblock {\em Archives of physical medicine and rehabilitation},
  88(1):99--104, 2007.

\bibitem{whitney2013symptoms}
S.~L. Whitney, P.~J. Sparto, J.~R. Cook, M.~S. Redfern, and J.~M. Furman.
\newblock Symptoms elicited in persons with vestibular dysfunction while
  performing gaze movements in optic flow environments.
\newblock {\em Journal of Vestibular Research}, 23(1):51--60, 2013.

\bibitem{zare2018developing}
M.~M. Zare, A.~Aslani, M.~Fakhrahmad, and S.~Ezzatzadegan.
\newblock Developing an android-based patient decision aid based on ottawa
  standards for patients after kidney transplant and its usability evaluation.
\newblock {\em Studies in health technology and informatics}, 249:61--68, 2018.

\end{thebibliography}
\end{document}